\documentstyle[12pt,epsf,epsfig]{article}
\begin{document}
\begin{flushright}
{\large UGI-97-09}
\end{flushright}
\begin{center}
{\Large \bf
The Spectral Function of the Rho Meson\\
in Nuclear Matter
\footnote{Work supported by BMBF and GSI Darmstadt.}}
\bigskip
\bigskip

{W. Peters$^1$, M. Post$^1$, H. Lenske$^1$, S. Leupold$^1$ and U. Mosel$^{1,2}$}

\bigskip
\bigskip
{ \it
$^1$Institut f\"ur Theoretische Physik, Universit\"at Giessen,\\
D-35392 Giessen, Germany\\
\bigskip
$^2$ Institute for Nuclear Theory, \\
University of Washington, Box 351550, Seattle, 98195, USA\\
}
\end{center}

\bigskip
\bigskip

\begin{abstract}
We calculate the modification of a rho meson in nuclear matter through
its coupling to resonance-hole states.  Starting from a recently proposed
model, we include all four star resonances up to 1.9 GeV. In contrast to
previous works, we include not only resonances that couple to the rho
in a relative $p$-wave, but also those that couple to an $s$-wave state.
In addition, we solve the equation for the rho spectral function
self-consistently.  We find that $s$-wave resonances affect the in
medium spectral function of the rho strongly.  
In the transverse channel 
the rho meson is, especially at non zero momentum, completely
washed out and can in the presence of nuclear matter no longer be
viewed as a resonant excitation of
the vacuum.  Instead, our model shows a
continuum of possible excitations with the quantum numbers of 
a transversely polarized rho. 
In the longitudinal channel, however, the rho retains its 
resonant character in our calculation.
As a consequence of the self-consistent treatment 
we also find a strong enhancement of the widths of the included
nucleon resonances in medium.
\end{abstract}

\section{Introduction}

There has been an ongoing discussion about the modification of the rho
meson in hadronic matter for several years, but up to now the problem
is not completely resolved.  Due to an argument of Brown and
Rho \cite{brownrho}, based on chiral symmetry arguments and scale
invariance, the mass of the rho meson should drop by about 15\% at
normal nuclear density. QCD sum rules arrive at a similar result, using,
however, a very simple spectral function of the rho \cite{hatsuda}
(For a more detailed discussion see also \cite{us}).
Since these works link the rho mass to the scalar quark condensate, a
dropping rho mass could be considered a signal for the restoration of
chiral symmetry.  Another argument by Pisarski \cite{pisarsky},
however, also using chiral symmetry arguments, predicts a mass that
increases as chiral symmetry is restored.

All these studies use rather general concepts based on first principle
arguments.  A complete understanding of these effects, however,
requires a reliable description of the modification of a rho meson in
nuclear matter due to purely hadronic interactions and up to now there
is no complete model available for this. Some works
\cite{herrmann,chanfray} consider the renormalization of the
$\pi\pi$ self-energy loop of the rho in medium using the delta-hole
model.  In \cite{asakawa} this approach is combined with both the
works
\cite{brownrho} and  \cite{hatsuda}, and the authors claim that 
Brown-Rho scaling has to be introduced in order 
to reach consistency with the
QCD sum rules. In \cite{weise}, however, consistency with QCD sum rules
is reached by using a chiral effective Lagrangian to calculate the rho
self-energy in a $T \rho_{_N}$ approximation without any further mass
modification. In \cite{eletsky} a similar approximation is used to calculate
a mass shift for higher three-momenta from the Compton-amplitude.

The CERES-data \cite{ceres1,ceres2,ceres3} caused a new discussion of this 
subject
since theoretical models \cite{cassing,ko} that used the vacuum
properties of the rho clearly underestimated the dilepton yield for
invariant masses in the range of about 300-600 MeV 
(see, however, \cite{koch}).
The agreement with the data could be improved, however, 
by either introducing Brown-Rho scaling for the rho
mass \cite{cassing,ko}, or by using the in-medium modification of the
$\pi\pi$ loop \cite{rapp1}.

Recently Friman and Pirner \cite{friman} calculated the contribution
of resonance-hole loops to the rho self-energy in nuclear
matter.  In \cite{rapp} this model was extended by including the
$\Delta(1232)$, the modification of the $\pi\pi$ loop in medium and
finite temperature effects.

In \cite{friman} only two (three in \cite{rapp}) resonances were
taken into account, all of which couple to the rho in a relative
$p$-wave. 
Since only $p$-wave
resonances were included in
\cite{friman}, it was found that the spectral function of the rho 
in vacuum was restored for vanishing three-momentum since the coupling
to these resonances vanishes in this limit.

We have extended the model presented in \cite{friman} by
including all four-star nucleon resonances up to $m^* \sim 1.9$ GeV
\cite{pdg}. 
Several of these resonances 
$(N(1520),\Delta (1620), N(1650),\Delta (1700))$
can couple to the rho and a nucleon in a
relative $s$-wave, so
that also a rho meson at rest will undergo modifications because of a
coupling to these channels.
In this paper we explore the consequences of these
new degrees of freedom and, in addition, present results of calculations in 
which the meson and nucleon resonance properties are treated 
self-consistently. 

In the next section we discuss the structure of the rho-propagator in medium.
In Sec.~\ref{theo} we present our model for the rho self-energy,
our results are shown in Sec.~\ref{results}. In Sec.~\ref{concl} we
summarize our results.


                        \section{The Rho-Propagator}

The propagator of a free, stable spin-1 particle with mass $m_o$ is given by:
\begin{eqnarray}
\label{freeprop}
D^{o \,\mu \nu}(q) &=& \frac{-(g^{\mu \nu}-\frac{q^{\mu}\,q^{\nu}}{m_o^2})}
                            {q^2-m_o^2} \nonumber\\
 &=& \frac{-(g^{\mu \nu}-\frac{q^{\mu}\,q^{\nu}}{q^2})}
                            {q^2-m_o^2} +
                                \frac{1}{m_o^2}\,\frac{q^{\mu}\,q^{\nu}}{q^2}
\quad ,
\end{eqnarray}
where in the second line of Eq.~(\ref{freeprop}) we have separated the
propagator into a longitudinal and a transverse part.  The coupling of
the rho to the pionic current gives rise to a self-energy
$\Sigma_{\pi\pi}^{\mu \nu}(q^2)$. Due to current conservation, the
self-energy is transverse:
\begin{equation}
\label{currcons}
        q_{\mu}\,\Sigma_{\pi\pi}^{\mu \nu}(q^2) = 0
\quad .
\end{equation}
Because of this, only the first term in Eq.~(\ref{freeprop}) is
modified, when a self-energy is taken into account:
\begin{eqnarray}
\label{vacprop}
D^{\mu \nu}(q)  &=& \frac{-(g^{\mu \nu}-\frac{q^{\mu}\,q^{\nu}}{q^2})}
                            {q^2-m_o^2 - \Sigma_{\pi\pi}(q^2)} +
                                \frac{1}{m_o^2}\,\frac{q^{\mu}\,q^{\nu}}{q^2}
\end{eqnarray}
with
\begin{eqnarray}
\Sigma_{\pi\pi}(q^2)=-\frac{1}{3}\,g_{\mu \nu}\,\Sigma_{\pi\pi}^{\mu \nu}(q^2)
\quad .
\end{eqnarray}
The real part of $\Sigma_{\pi\pi}$ is taken into account approximately
 by putting $m_o
\equiv m_\rho$, where $m_\rho$ is the physical rho-mass. 
Thus we take:
\begin{eqnarray}
\Sigma_{\pi\pi}(q) \equiv - {\rm i} \;m_\rho \;\Gamma_{\pi\pi}(q^2)
\quad .
\end{eqnarray} 
$ \Gamma_{\pi\pi}$ is the decay width of the rho into two pions. We use the
parameterization:
\begin{eqnarray}
\label{rhowidth}
        \Gamma_{\pi \pi}(m^2) = \Gamma_o \,\frac{m_\rho}{m}\, 
        \left(\frac{q\left(m\right)}{q\left(m_\rho \right)}\right)^3 \,\,.
\end{eqnarray}
$q(m)$ is the momentum of the pions in the restframe of the decaying
rho having mass $m$; $m_\rho$ is the physical rho-mass and $\Gamma_o$
the corresponding width.  Eq.~(\ref{rhowidth}) contains precisely the
energy dependence of the imaginary part of the self-energy that comes
out of a one loop calculation. Since this reproduces
 the $\pi\pi$-scattering data up to 
$\sqrt{s}\sim 1$ GeV
\cite{weise0}, we do not include an additional formfactor here,
which would affect the rho width only in the region of very high
invariant mass.
The energy dependence of the real part of the $\pi\pi$-loop has
virtually no influence on the vacuum spectral function defined as:
\begin{eqnarray}
\label{freespec}
A(q^2) = -\frac{1}{\pi} \; {\rm Im}\left(
        \frac{1}{q^2 - m_\rho^2 -\Sigma_{\pi\pi}}\right)
\quad 
\end{eqnarray}
and is therefore not included in our calculation.

In medium Lorentz-invariance is broken and the rho self-energy is
characterized by two scalar functions instead of one. Using the
longitudinal and transverse projectors \cite{weise}:
\begin{eqnarray}
\label{projectorst}
P_T^{\mu\nu} &=& - \left(
        \begin{array}{cc}
                0&0\\
                0&\delta_{ij} - \frac{q_iq_j}{\vec{q}\,^2}\\
                \end{array} 
                \right)
\end{eqnarray}
and
\begin{eqnarray}
\label{projectorsl}
P_L^{\mu\nu} &=& (g^{\mu \nu}-\frac{q^{\mu}\,q^{\nu}}{q^2}) - P_T^{\mu\nu}
\nonumber \\
&=& -\left(
        \begin{array}{cc}
          \frac{\vec{q}\,^2}{q^2}&\frac{\omega q_j}{q^2}\\
          \frac{\omega q_i}{q^2}&\frac{ \omega^2 q_iq_j}{\vec{q}\,^2 q^2}\\
                \end{array} 
                \right)
\end{eqnarray}
we can decompose the self-energy of the rho in medium:
\begin{eqnarray}
        \Sigma^{\mu\nu}(q)= -
                P_L^{\mu\nu}\,\Sigma^L(q) - 
                P_T^{\mu\nu}\,\Sigma^T(q)  
\end{eqnarray}
with the longitudinal and transverse parts
\begin{eqnarray}
\begin{array}{rcrl}
        \Sigma^L(\omega,\vec{q}) &=&
                -& P_L^{\mu\nu}\Sigma_{\mu\nu}(\omega,\vec{q}) 
\\
\\
        \Sigma^T(\omega,\vec{q}) &=& -\frac{1}{2} &
                P_T^{\mu\nu}\Sigma_{\mu\nu}(\omega,\vec{q})
\quad .
\end{array}
\end{eqnarray}
Then the dressed rho propagator becomes:
\begin{eqnarray}
        D^{\mu\nu}(\omega,\vec{q})=-\frac{P_L^{\mu\nu}} 
                {q^2-m_\rho^2-\Sigma^L(\omega,\vec{q})}
                -\frac{P_T^{\mu\nu}}{q^2-m_\rho^2-\Sigma^T(\omega,\vec{q})}
                +\frac{q^\mu q^\nu}{m_\rho^2 q^2}
\quad .
\end{eqnarray}
Thus a transverse and a longitudinal spectral function can be defined:
\begin{eqnarray}
\label{fullspec}
A^{T(L)}(\omega,\vec{q}) = - \frac{1}{\pi} \;{\rm Im}\left(
        \frac{1}{q^2 - m_\rho^2 -\Sigma^{T(L)}(\omega,\vec{q})
        -\Sigma_{\pi\pi}(q^2)}
\right) \quad .
\end{eqnarray}
For $\vec{q}=0$,  $A^T=A^L$ is obtained.

        \section {The Rho Self-Energy in Matter}
\label{theo}

Similar to the pion, that becomes renormalized by delta-hole loops in
the nuclear medium, there are contributions to the self-energy of
the rho by resonance-hole loops. Since we are interested in a broad
range of energies, there will be several resonances contributing. We
include all four-star resonances up to an energy of 1.905 GeV.
Only the $N(1440)$, for which the partial decay width
into a rho and a nucleon is poorly known and compatible with zero
\cite{pdg,manley}, is excluded. The resonances we include are listed in Table
\ref{listres} together with their quantum numbers.

As in \cite{friman,rapp} we calculate 
the loop contributions to the self-energy of the rho in nuclear matter
in a
non-relativistic approximation. The non-relativistic coupling terms
are constructed by reduction of relativistic interaction Lagrangians
that are compatible with parity conservation and gauge
invariance. Depending on the quantum numbers of the resonance we
arrive at the following interactions:
\begin{eqnarray}
\label{couplings}
\begin{array}{llllll}                                           
{\cal L}_{int}&=&       \frac{f_{R N\rho}}{m_\rho}\,\,
                \psi_{R}^\dagger\,\sigma_k\,\epsilon_{ijk}\,q_i\,\rho_j\,\psi
                &\mbox{for}&   \frac{1}{2}^+                            \\ \\
&=&     \frac{f_{R N\rho}}{m_\rho}\,\,
                \psi_{R}^\dagger\,S_k\,\epsilon_{ijk}\,q_i\,\rho_j\,\psi
                &\mbox{for}&   \frac{3}{2}^+                            \\ \\
&=&     \frac{f_{R N\rho}}{m_\rho}\,\,
                \psi_{R}^\dagger\,(\sigma_k\,\rho_k\,\omega
                - \rho_o\sigma_k\,q_k) \,\psi
                &\mbox{for}&   \frac{1}{2}^-                            \\ \\
&=&     \frac{f_{R N\rho}}{m_\rho}\,\,
                \psi_{R}^\dagger\,(S_k\,\rho_k\,\omega
                - \rho_o S_k\, q_k )\,\psi
                &\mbox{for}&   \frac{3}{2}^-                            
\quad .
\end{array}
\end{eqnarray}
$S_k$ is the spin-$\frac{3}{2}$ transition operator.
The isospin parts of the coupling terms are:
\begin{eqnarray}
\begin{array}{lllll}
 & \chi_{R}^\dagger\,\sigma_k\,\rho_k\,\chi &\mbox{for}&
 I=\frac{1}{2} &\\ 
 &      \chi_{R}^\dagger\,S_k\,\rho_k\,\chi
                &\mbox{for}& I=\frac{3}{2} &
\quad .
\end{array}
\end{eqnarray}
$\chi_{R}$, $\chi$ and $\rho$ stand for the isospin part of the 
wavefunction of the resonance, the
nucleon and the rho, respectively.

All these couplings are current conserving, i.e. they yield zero under
the replacement $\rho_\mu \to q_\mu=(\omega,\vec{q})$. This has to be
the case since the rho self-energy has to fulfill the condition of
current conservation, Eq.~(\ref{currcons}), in medium as well. For the
$\Delta(1905)$ with $J^P = \frac{5}{2}^+$ we take:
\begin{eqnarray}
\label{couplings52}
\begin{array}{llllll}                                           
{\cal L}_{int}&=&       \frac{f_{R N\rho}}{m_\rho}\,\,
                \psi_{R}^\dagger\,R_{ij}\,q_i\,\rho_j^{T}\,\psi
                && \\ \\
\end{array}
\quad .
\end{eqnarray}
In this equation $R_{ij}$ is the spin-$\frac{5}{2}$ transition operator 
\cite{friman}
and $\rho_j^{T}$ stands for the transverse projection of
the rho polarization vector:
\begin{eqnarray}
\rho_j^{T} \;\; = \;\; \left( \delta_{jk} - \frac{q_j q_k}{\vec{q}\,^2} \right)
\rho_k
\quad .
\end{eqnarray}
This projection is necessary to ensure current conservation.
In \cite{friman} this coupling is given without the
projection onto transverse polarizations, but since only the
transverse spectral function is calculated there, it is effectively taken
into account.

Note that those resonances decaying into a rho in a relative $p$-wave
have a coupling proportional to $\vec{q}$, the three momentum of the
rho, while the ones coupling to an $s$-wave state 
contain $\vec{q}$ and 
$\omega$, the energy of the rho.  In the case $\vec{q} = 0$ only the
latter will contribute. Since in \cite{friman,rapp} only $p$-wave
resonances are taken into account, in these works the vacuum propagator
is recovered for $\vec{q} = 0$. If $s$-wave resonances are included, this
is no longer the case (see sec. \ref{results}).

\begin{figure}
\begin{center} 
\epsfig{file=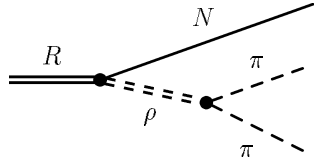,width=3.5cm}
\end{center}
\caption{\label{rdecaygraph}
Decay of a nucleon resonance into two pions via a rho meson.}
\end{figure}

The coupling constants $f_{R N\rho}$ are determined by calculating the
partial decay width of each resonance for a decay into a nucleon and
two pions via the rho (see Fig.~\ref{rdecaygraph}):
\begin{eqnarray}
\label{rdecay1}
        \begin{array}{ll}
        \Gamma(m_{_R}) &= \left(\frac{f_{R N\rho}}{m_\rho}\right)^2
        \frac{1}{\pi}
        \,S_{\Gamma} \;\;  \\
        \\
        &\times\;\;{\displaystyle \int \limits_{{2m_\pi}}
        ^{m_{_R}-m_{_N}}}\!\!\!\!\!\!dm\,\,
        m\,\,\frac{m_{N}}{m_{_R}}\,\,|\vec{q}\,|^3\,\,
        A(m^2)\,\,F(\vec{q}\,^2)
        \end{array}
\end{eqnarray}
for $p$-wave decays and
\begin{eqnarray} 
\label{rdecay2}
        \begin{array}{ll}
        \Gamma(m_{_R}) &= \left(\frac{f_{R N\rho}}{m_\rho}\right)^2
        \frac{1}{\pi}
        \,S_{\Gamma}\;\; \\
        \\
        &\times\;\;{\displaystyle \int \limits_{2m_\pi}
        ^{m_{_R}-m_{_N}}}\!\!\!\!\!\!dm\,
        m\,\,\frac{m_{N}}{m_{_R}}\,\,|\vec{q}\,|
        \,(2\omega^2+m^2) \,\, 
        A(m^2)
        \,\,F(\vec{q}\,^2)
        \end{array}
\end{eqnarray}
for $s$-wave decays.

$q(m)$ denotes the three-momentum of the rho in the center of mass
frame, $\omega$ its energy and $m$ its invariant mass. 
$m_{_N}$($m_{_R}$) is the mass of the nucleon (the
resonance). $S_{\Gamma}$ stands for the factor arising from
spin/isospin summation, its values are also given in Table
\ref{listres}. 
$A(m^2)$ is the vacuum spectral function
from Eq. (\ref{freespec}) and $F(\vec{q}\,^2)$ is a monopole
formfactor:
\begin{equation} 
        F(\vec{q}\,^2) = \frac{\Lambda^2}{\Lambda^2+\vec{q}\,^2} \quad
        \mbox{with} \quad \Lambda = 1.5\,{\rm GeV}.
\end{equation} 
The coupling constants $f_{R N\rho}$ are then adjusted to reproduce
the partial decay widths \cite{pdg};
together with these widths they are given in Table 1.
Since the $\Delta(1232)$ does not show a
measurable branching ratio into two pions, its coupling to the rho
cannot be determined by calculating a decay width.
Instead, we take the value also used
in \cite{rapp}. The $NN\rho$ coupling constant is also taken to be the same
as in \cite{rapp}.

For invariant masses of the resonances above their free
mass, the rho-widths can become very large. This is mainly due to
phase space, that becomes larger for increasing invariant mass of the
resonance.
To avoid unphysical values, we cut them off at 1  GeV.
The precise value of this cutoff has only very little influence
on our results shown in sec. \ref{results}.

\begin{figure}
\begin{center} 
\epsfig{file=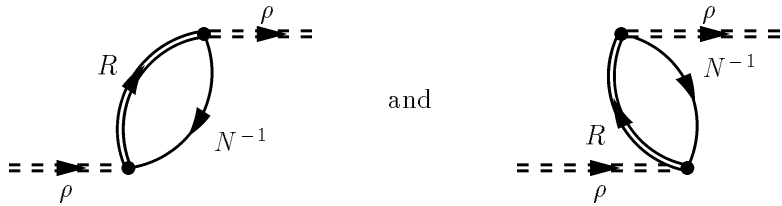,width=10cm}
\end{center}
 \caption{\label{loops}
Resonance-hole self-energy diagrams for the rho meson. The double dashed line 
stands for a physical rho meson that contains the $\pi\pi$-width.}
 \end{figure}

Now we have all the ingredients necessary to calculate the resonance-hole loops
given in Fig.~\ref{loops}: 

\paragraph{$p$-wave resonances:}

Evaluating the diagrams in Fig.~\ref{loops} we find for resonances
with positive parity:
\begin{equation}
\label{selfpos}
        \Sigma^{ij}(\omega,{\vec{q}}) =
        \left(\delta^{ij}-\frac{q^i\,q^j}{\vec{q}\,^2}\right)\,
        \left(\frac{f_{R
        N\rho}}{m_\rho}\right)^2\,\vec{q}\,^2\,F(\vec{q})
        \,\,\,S_{\Sigma}\,\,\,\beta(\omega,{\vec{q}})
        \\
\end{equation}                  
and
\begin{equation}
\Sigma^{\mu 0}\; = \;\Sigma^{0 \nu}\; = \; 0 
\end{equation}                  
$\Sigma^{\mu\nu}$ as given in Eq.~(\ref{selfpos}) is obviously current
conserving and purely transverse.  The spin-isospin factors
$S_{\Sigma}$ are also given in Table \ref{listres}.  The function
$\beta(\omega,{\vec{q}})$ is given by:
\begin{eqnarray}
\label{beta}
        \begin{array}{lll}
        &\beta(\omega,\vec{q})=&  \\
        \\
        &{\displaystyle
        -\int\limits_0^{p_F} \!\! \frac{d^3\!p}{(2\,\pi)^3}\,\,
        \left(\frac{1}{\omega-E_{_N}({\vec{p}})+E_{_R}(\vec{p}+\vec{q})}+
        \frac{1}{-\omega-E_{_N}(\vec{p})+E_{_R}(\vec{p}-\vec{q})}\right) \,\,}&
        \end{array}
\end{eqnarray}
In the limit of small nuclear density $\rho_{_N}$ or large meson momenta
($|\vec{q}\,| \gg p_{F}$),
$\beta(\omega,\vec{q})$ reduces to:
\begin{equation}
\label{lowdens}
        \beta(\omega,\vec{q})= \frac{1}{2}\,\rho_{_N}\,
        \frac{E_{_R}(\vec{q})-m_{_{N}}}
                {\omega^2-(E_{_R}(\vec{q})-m_{_{N}})^2}
\quad .
\end{equation}
$E_{_N}(\vec{q})$ and 
$E_{_R}(\vec{q})$ are given by:
\begin{eqnarray}
        \begin{array}{rcl}
        E_{_N}(\vec{q}) &=& \sqrt{m_{_{N}}^2\!+\vec{q}\,^2 }\\ \\
        E_{_R}(\vec{q}) &=& \sqrt{m_{_R}^2\,\,+\vec{q}\,^2} 
                - \frac{{\rm i}}{2}\,\Gamma_{_{R}} \,\,\,\,.
        \end{array}
\quad 
\end{eqnarray}
$\Gamma_{_{R}}$ is the total width of the resonance, it is taken to be
the sum of the width for pion and rho decay:
\begin{equation}
\label{totwidth}
        \Gamma_{_R}=\Gamma_{_{R\rightarrow N\pi}}+
        \Gamma_{_{R\rightarrow N\rho}}
\quad .
\end{equation}
$\Gamma_{_{R\rightarrow N\rho}}$ is calculated according to 
Eq.~(\ref{rdecay1}), 
$\Gamma_{_{R\rightarrow N\pi}}$ is taken to be
\begin{equation}
\label{piwidth}
        \Gamma_{_{R\rightarrow N\pi}}=\Gamma_o\,
                \left(\frac{q}{q_r}\right)^{2\ell+1} \quad 
\quad .
\end{equation}
$q$ and
$q_r$ are the three momenta of a pion for a resonance of
given mass and on its mass shell, respectively. $\ell$ is the orbital
angular momentum of this pion. 
The values for $\Gamma_o$ are given in Table \ref{listres}; 
they are adjusted such that $\Gamma_R$ from Eq. 
(\ref{totwidth}) is equal to the total empirical width.
We find that the actual momentum
dependence in Eq.~(\ref{piwidth}) has only very little influence on our
results (see sec. \ref{results}).

\paragraph{$s$-wave resonances:}

For the case of 
$s$-waves, i.e.  negative parity states,  one gets:
\begin{equation}
\label{selfneg}
        \Sigma^{\mu\nu}(\omega,\vec{q}) = 
        T^{\mu\nu}
        \left(\frac{f_{R N\rho}}{m_\rho}\right)^2\,
        F(\vec{q}\,^2)\,\,\,
        S_{\Sigma}\,\,\,\beta(\omega,\vec{q}) \,\, 
\quad .
\end{equation}  
This expression is evaluated in the same way
as for $p$-wave resonances. The tensor $T^{\mu\nu}$ is given by
\begin{equation}
T^{\mu\nu}=
        \left(
          \begin{array}{cc}
            {\vec{q}\,^2}&{\omega q_j}\\
            {\omega q_i}&\omega^2 \delta_{ij}\\
          \end{array} 
        \right) 
\quad ,
\end{equation}  
and can be expressed in terms of the projectors from Eqs.~(\ref{projectorst})
and (\ref{projectorsl}):
\begin{equation}
\label{decs}
T^{\mu\nu}= -\omega^2 P_T^{\mu\nu} - q^2 P_L^{\mu\nu}
\quad .
\end{equation}  
It follows, that the contribution of $s$-wave
resonances to $\Sigma^T$ is proportional to $\omega^2$, while their 
contribution to $\Sigma^L$ goes like $q^2=\omega^2 - \vec{q}\,^2$. For
$\vec{q}=0$, this ensures that $\Sigma^T=\Sigma^L$, since the contribution
from $p$-wave resonances vanishes in this case (Eq. (\ref{selfpos})).
Thus, a longitudinally polarized rho will be modified only by $s$-wave
resonances, while a transverse rho couples to both $s$- and $p$-wave
resonances.

\paragraph{}

In \cite{rapp} Landau-Migdal parameters are introduced in the
expressions (\ref{selfpos}) to account for additional short-range
correlations.
The value of these parameters for
resonances beyond the $\Delta(1232)$ can, however, only be guessed.
Anyway, we find that the dependence of our results on the inclusion of
Landau-Migdal parameters for $p$-wave resonances beyond the
$\Delta(1232)$ is negligible.  For the $\Delta(1232)$ we take:
\begin{equation}
\Sigma_{\Delta} (\omega,\vec{q})\; = \;
        \frac { \Sigma^{o}_{\Delta} }{ 1- g'
                \frac{\Sigma^{o}_{\Delta}}{\vec{q}\,^2}       }
\end{equation}
with $g'=0.5$.

The spectral function in Eq.~(\ref{fullspec}) contains the rho-width
of the resonances, which in turn contains the spectral function via
Eq.~(\ref{rdecay1}) and (\ref{rdecay2}).  This implies a change of the rho
decay-width of a resonance in medium.  We solve this problem
self-consistently by iterating Eq.  (\ref{fullspec}), starting with
the vacuum spectral function. The first step of this iteration is
equivalent to the calculations in
\cite{friman}. 

The resonance widths in medium are given in terms of the two 
spectral functions $A^T$ and $A^L$ by:

\begin{eqnarray}
\label{rdecayim1}
        \begin{array}{ll}
        \Gamma(m_{_R}) &= \left(\frac{f_{R N\rho}}{m_\rho}\right)^2
        \frac{1}{2\pi}
        \,S_{\Gamma} \;\;  {\displaystyle \int 
        }q\;dq\;d\!\cos \theta\\
        \\
        &\times\;\;
        \,\,\frac{m_{N}}{E_N}\,\,|\vec{q}\,|^3\,\,
        A^T(\omega,\vec{q})\,\,F(\vec{q}\,^2)( 1-n_f(\vec{q})) 
        \end{array}
\end{eqnarray}
for $p$-wave resonances and
\begin{eqnarray} 
\label{rdecayim2}
        \begin{array}{ll}
        \Gamma(m_{_R}) &= \left(\frac{f_{R N\rho}}{m_\rho}\right)^2
        \frac{1}{2\pi}
        \,S_{\Gamma}\;\; {\displaystyle \int 
        }q\;dq\;d\!\cos \theta\\
        \\
        &\times\;\;
        \,\,\frac{m_{N}}{E_N}\,\,|\vec{q}\,|
        \,(2\omega^2A^T(\omega,\vec{q})+m^2A^L(\omega,\vec{q})) \,\, 
        \,\,F(\vec{q}\,^2)( 1-n_f(\vec{q}))
        \end{array}
\end{eqnarray}
for $s$-wave resonances. $\cos \theta$ is the emission angle of the
rho in the cm-frame of the decaying resonance and $q$ is its
three-momentum. In the vacuum, i.e. for $A^L=A^T$, this reduces to the
expressions (\ref{rdecay1}) and (\ref{rdecay2}).

                        \section{Results and Discussions}
\label{results}

Within the framework described in the previous sections we have
calculated the spectral functions $A^T$ and $A^L$ of the rho meson in
nuclear matter.

In order to make the interpretation of the following figures easier,
we show in Fig.~\ref{branches} the free dispersion relations for the
rho, the $\Delta(1232)$, the $N(1520)$ and the $N(1720)$, since these
are the resonances that can be identified most easily in the following
plots. For the resonances $m$ stands for the invariant mass
that a meson of given momentum must have in 
a collision with a nucleon at rest in order to excite the
respective resonance on its mass shell.

\begin{figure} 
\epsfig{file=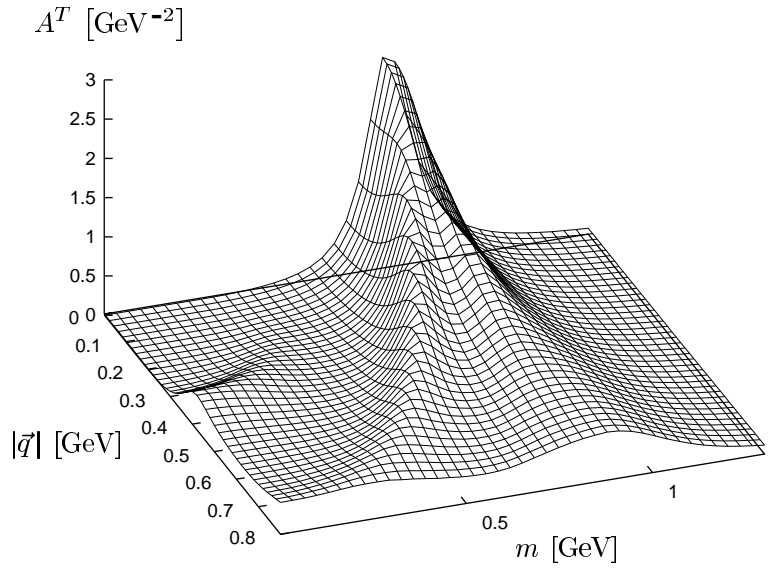,width=10cm}
 \caption{
\label{ponly1t}
Non-selfconsistent transverse spectral function of the rho
 meson in nuclear matter as defined in the text for 
$\rho_{_N} =  \rho_o$ as a function of invariant mass 
$m$ and three-momentum $q$. 
Only $p$-wave resonances are included. }
\end{figure}

In  Fig.~\ref{ponly1t} we show $A^T$ as a function
of three-momentum and invariant mass of the rho when only $p$-wave
resonances are included, as the result of a non self-consistent
calculation.  This spectral function can be directly compared to that
obtained by Friman and Pirner \cite{friman}. Even though these authors
include only two resonances, their result is qualitatively
 similar to the one
shown in Fig.~\ref{ponly1t}.  At $\vec{q}=0$ the
spectral function is equal to the one in vacuum, while for higher momenta
the rho-peak vanishes and a new structure appears at lower invariant
masses, which stems from the $\Delta(1232)$ and the $N(1720)$.
It must be kept in mind, that the value of the $\Delta N \rho$
coupling constant is rather uncertain, so that only qualitative 
conclusions can be made about its contribution.

\begin{figure} 
\epsfig{file=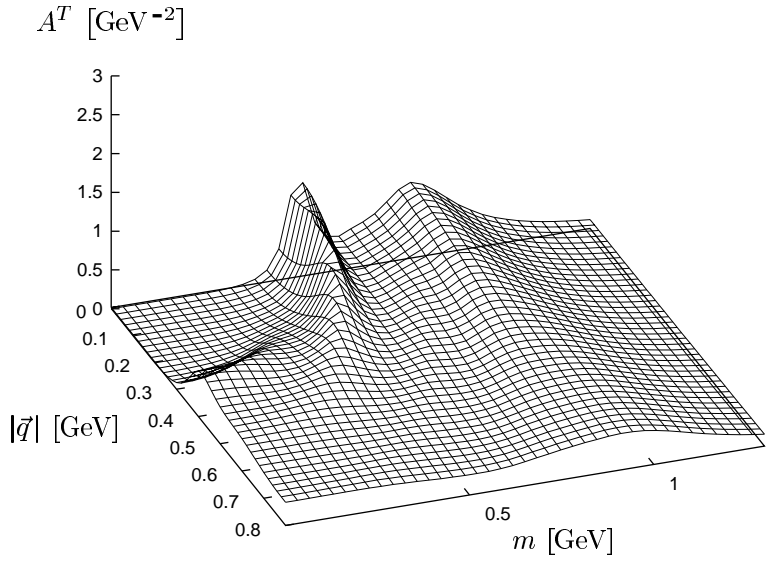,width=10cm}
\epsfig{file=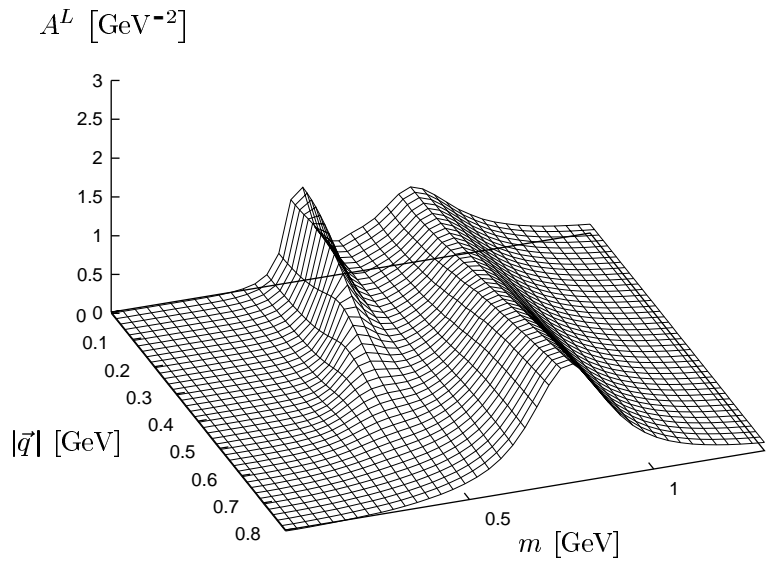,width=10cm}
 \caption{
\label{nonsc}
Non-selfconsistent spectral functions of the rho
 meson for $\rho_{_N} =  \rho_o$. 
Upper part: transverse spectral function. 
Lower part: longitudinal spectral function.}
\end{figure}

Including also the $s$-wave resonances leads to a drastic change
of $A^T$,
especially at small $\vec{q}$ (upper part of Fig.~\ref{nonsc}). The
rho-peak is strongly depleted 
also for small momenta and
a pronounced, sharp new ridge due to the $N(1520)$ appears at lower
invariant masses and momenta. The dominance of this branch is due to the
relatively large observed partial decay width of the $N(1520)$ of about 25
MeV into a rho and a nucleon \cite{pdg}. Since the resonance mass lies far below
the energetic threshold for this decay calculated from the peak mass of
the rho, this decay can proceed only through the tails of the rho-mass
distribution leading to a relatively large coupling constant.

In the lower part of Fig.~\ref{nonsc} we show the 
corresponding longitudinal
spectral function $A^L$. 
Again, the $N(1520)$ leads to a pronounced structure at
small $\vec{q}$. In contrast to $A^T$ in the upper part of this
Figure, the $N(1520)$-branch dies out for increasing $\vec{q}$. The
rho-branch is reappearing at higher invariant mass, but its peak is
reduced by a factor of two relative to the vacuum case. The vanishing
of the $N(1520)$-branch is due to the fact, that the contribution of
s-wave resonances to $\Sigma^L$ is proportional to $q^2 \equiv m^2$
(see Eq.~\ref{decs}), which decreases along the $N(1520)$-branch as
shown in Fig.~\ref{branches}.  This $m^2$-dependence, together with
the fact, that the rho-branch and the resonance branches become the
more separated, the higher the momentum is (see Fig.~\ref{branches}),
is also the reason why the rho-branch reappears in the longitudinal
channel. This is in contrast to the contribution of $s$- and $p$-wave
resonances to $\Sigma^T$, that goes with $\omega^2$ and $\vec{q}\,^2$,
respectively. Both these factors increase along a line of
constant $m$ in Fig.~\ref{nonsc}, and lead therefore to the very broad
distribution of strength for large $\vec{q}$ in the case of $A^T$.

Using the coupling given in Eq.~(\ref{couplings}) for $\frac{1}{2}^+$
resonances, we find that an $(N N ^{-1})$ loop yields only a
negligible contribution to the rho self-energy.

In all the calculations shown, we integrate over the Fermi-sphere of the
nucleons in medium instead of using the approximation
(\ref{lowdens}). We find, however, that this has only very little effect
on the final result. Using Eq.~(\ref{lowdens}) would lead to slightly
sharper resonance peaks in  Fig.~\ref{nonsc}, but would not
change any of the dominant features.
This shows that our calculation so far 
amounts essentially to taking into
account the linear term in an expansion of the self-energy into powers
of the nuclear density $\rho_{_N}$, which is equivalent to considering
reactions of the rho with only a single nucleon. This is a reasonable
assumption as long as the three-momentum is high enough, so that the
rho can actually resolve a single nucleon. For vanishing three momentum,
however, this is not the case, so that terms of higher order in the
density, i.e. reactions with more than one nucleon, will become
important.  These are contained in a self-consistent solution that
includes terms to an arbitrary order in the density, or, phrased
differently, reactions of the rho with an arbitrary number of
nucleons.

\begin{figure} 
\epsfig{file=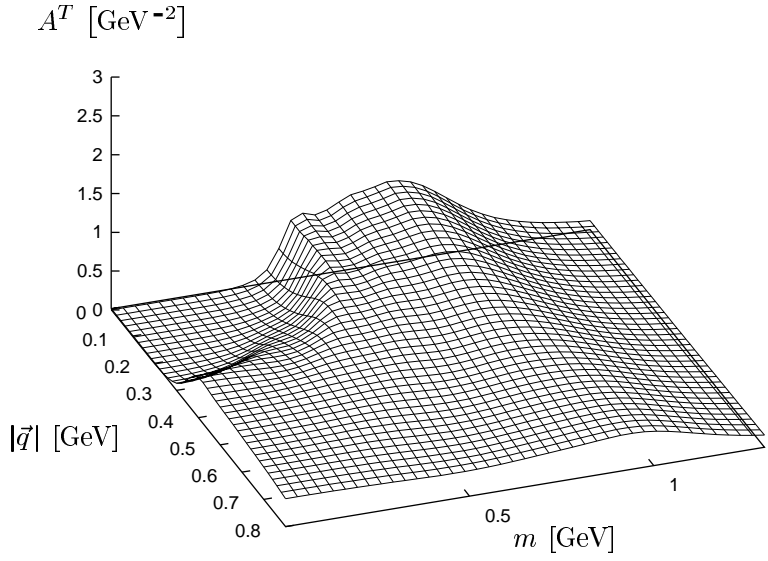,width=10cm}
\epsfig{file=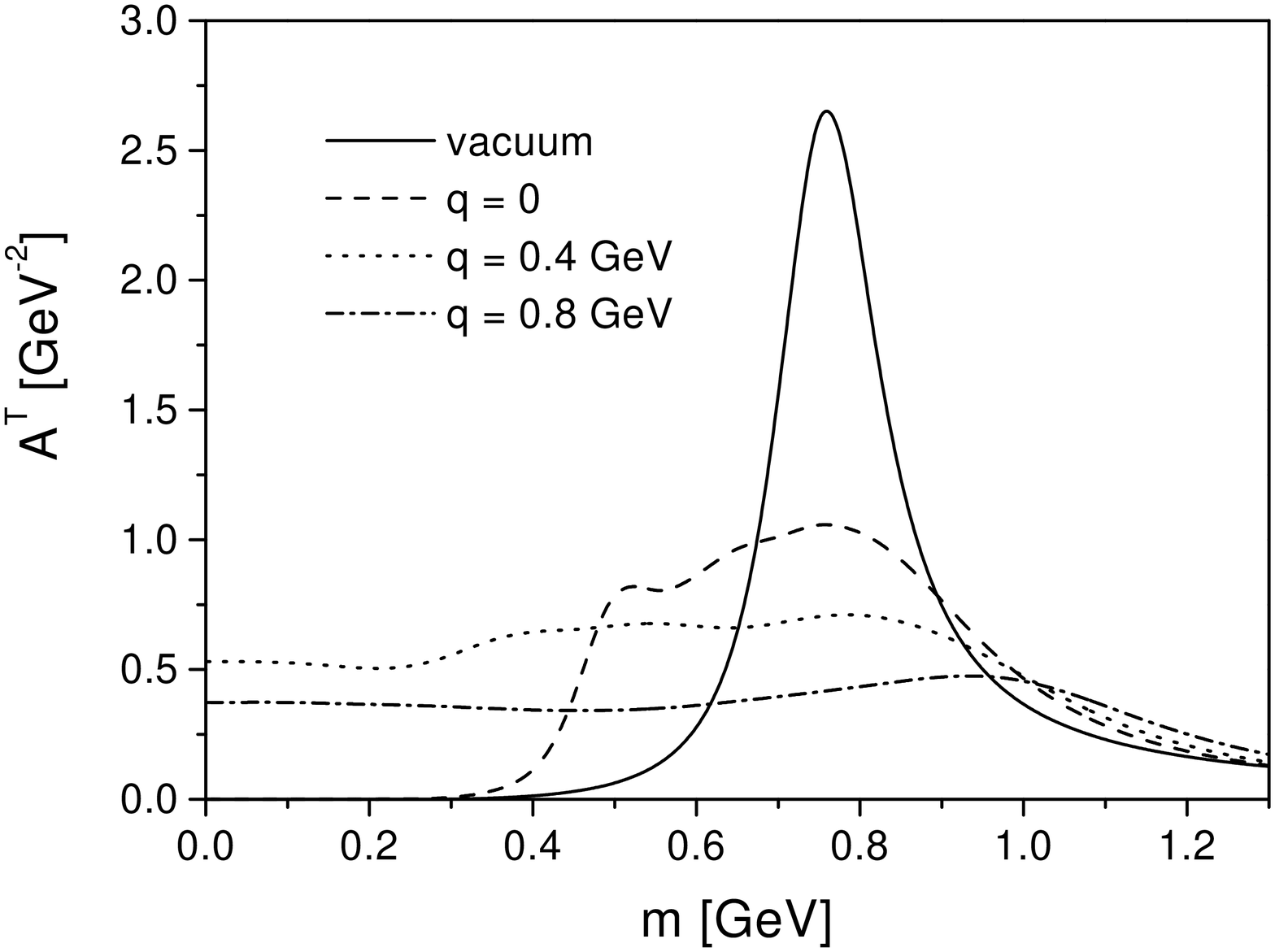,width=10cm}
 \caption{
\label{sct}
Self-consistent transverse spectral function of the rho
 meson for $\rho_{_N} =  \rho_o$.
Lower part: Cuts through the upper part for different three-momenta
together with the vacuum spectral function.}
\end{figure}

\begin{figure} 
\epsfig{file=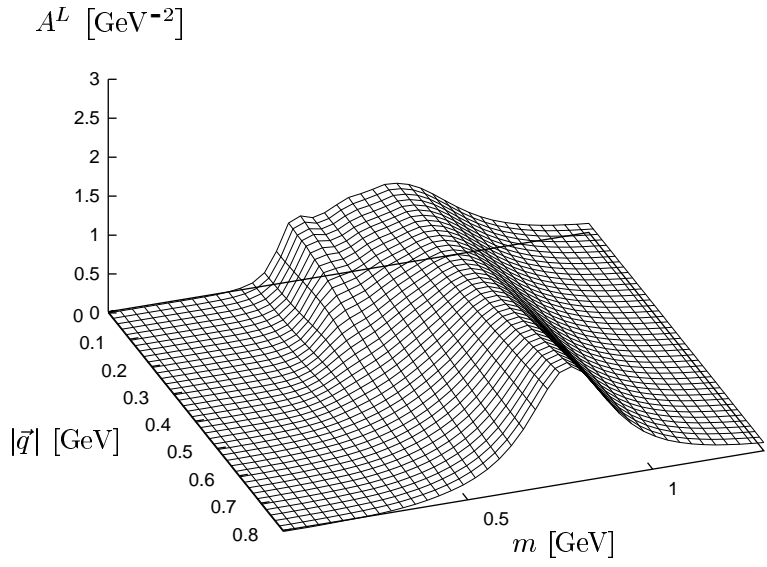,width=10cm}
\epsfig{file=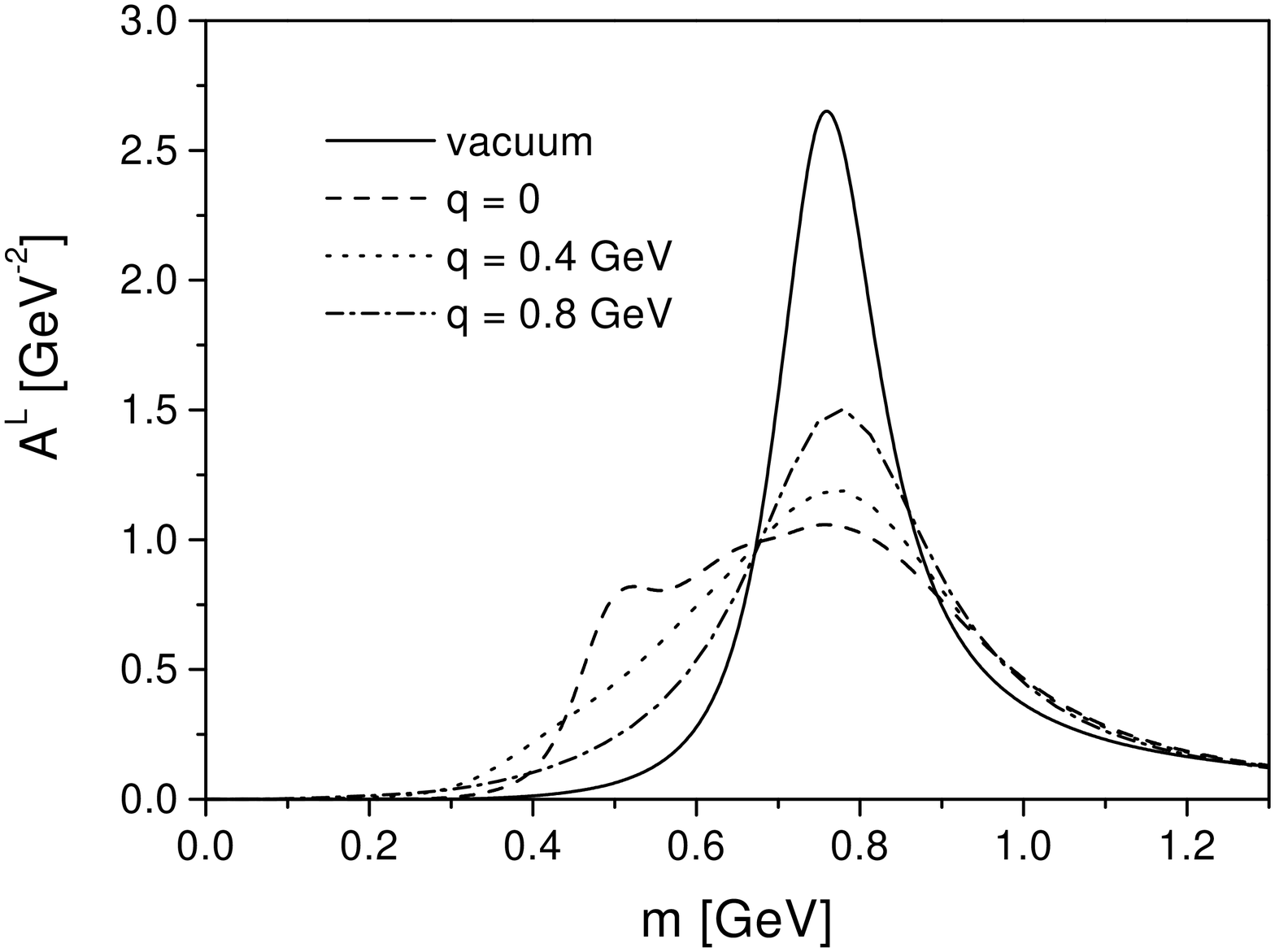,width=10cm}
 \caption{
\label{scl}
Same as Fig.~\ref{sct}, but for the longitudinal spectral function.}
\end{figure}

The result for such a self-consistent calculation is shown in
Figs.~\ref{sct} and \ref{scl} for normal nuclear density. For higher
three-momenta there are only small changes in both $A^T$ and $A^L$, 
while for $\vec{q}\to 0$
there is a strong influence of the terms of higher order in the
density. The main effect is that the structures visible in 
 Fig.~\ref{nonsc} at small $\vec{q}$ are washed out.

In the lower parts of Figs.~\ref{sct} and \ref{scl} we show cuts
through the upper parts of these figures for different three momenta
together with the vacuum spectral function.  Note that in
Fig.~\ref{sct}, in particular for momenta $|\vec{q}\,|>0.2$ GeV,
the transverse spectral function does not exhibit a resonance-like
structure, so that a description of the in-medium rho in terms of mass
and width is no longer appropriate for transverse polarization.
In contrast to that, the longitudinal spectral function still
shows a resonant behavior, especially for $|\vec{q}\,|>0.4$.

Thus we find, that at finite $\vec{q}$ a transverse rho undergoes a
stronger modification than a longitudinal one. The reason is that
only $s$-wave resonances contribute to the longitudinal channel, and
their effect decreases with increasing $\vec{q}$. The transverse
channel on the other side, is also modified  by $p$-wave resonances,
whose contributions to the rho self-energy increase (see above).
This difference between the polarizations agrees qualitatively
with the findings of QCD sum rules at finite $\vec{q}$ \cite{lee}, as
well as the analysis in \cite{eletsky}.

\begin{figure} 
\epsfig{file=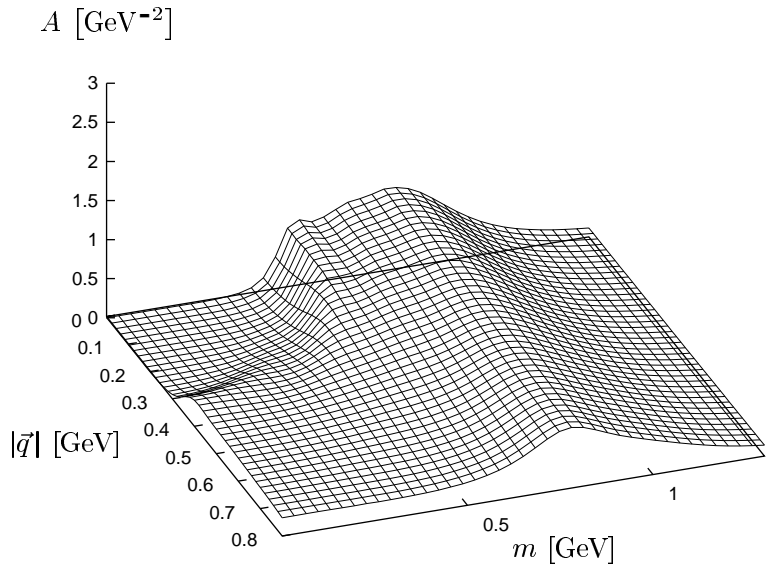,width=10cm}
\epsfig{file=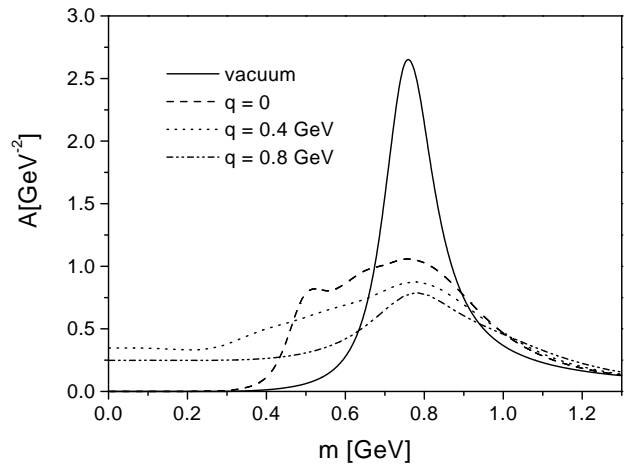,width=10cm}
 \caption{
\label{sca}
Same as Fig.~\ref{sct}, but for the averaged spectral function.}
\end{figure}

While these different polarizations of the rho could be exploited in
experiments using polarized particles, e.g.  real, transverse photons, in
the incoming beam, for the description of heavy-ion collisions only the
statistically averaged spectral function
$A = \frac{1}{3}(2A^T+A^L)$ is relevant. In Fig.~\ref{sca} we show $A$ in the
same way as $A^L$ and $A^T$ in Figs.~\ref{sct} and \ref{scl}.
At $\vec{q}=0$, we have $A = A^L = A^T$; for higher momentum
a maximum coming from the longitudinal rho is still visible, but
it sits upon a strong transverse background.

The self-consistent calculation has another surprising consequence,
namely that, except for the $\Delta(1232)$, the rho decay-widths of
the resonances increase dramatically. This is easy to understand,
since there is more strength in the region of small invariant masses
of the rho, where the phase space for the decay becomes larger.  The
values of these partial widths in vacuum for on-shell resonances are
given in Table
\ref{listwidths}, together with the values resulting from a
non-selfconsistent calculation, as well as a self-consistent
calculation at two different nuclear densities.  The difference
between the free value and the result of the non-selfconsistent
calculation is due to Pauli-blocking in the medium.  The widths for
$\rho_N = 2 \rho_o$ are smaller than  for normal nuclear
density due to the increasing importance of Pauli-blocking.  In
addition, one sees from Fig.~\ref{branches}, that the branches of
$\Delta(1232)$ and $N(1520)$ dive down into the region with $m^2<0$,
so that the overall amount of transverse strength in the region
relevant for the resonance decay ($m^2>4 m_{\pi}^2$), decreases.

The results in
Table \ref{listwidths} are for resonances with $\vec{q}=0$; we find
only a moderate dependence of these widths on the three-momentum.
The $\Delta(1232)$ is practically left unchanged, its rho-width 
stays below 1 MeV. 

As already discussed in \cite{effe} the significant broadening of the nucleon
resonances with large rho-decay widths also contributes to the observed
disappearance of the nucleon resonances beyond the Delta in the total
photoabsorption cross sections on nuclei \cite{bianchi}. 
While Fermi-motion alone
already leads to a significant broadening of the absorption cross section
on the individual, in particular the higher-lying, resonances, it is not
sufficient to explain the disappearance of the $N(1520)$. This resonance
which has a large electromagnetic strength is just the one that is
broadened the most in our calculation. Also other effects, such as the
onset of the 2$\pi$ channel, contribute most probably to the
disappearance of the $N(1520)$ in the photoabsorption cross section; a
quantitative evaluation of this mechanism, however, so far does not exist. 

\begin{figure} 
\epsfig{file=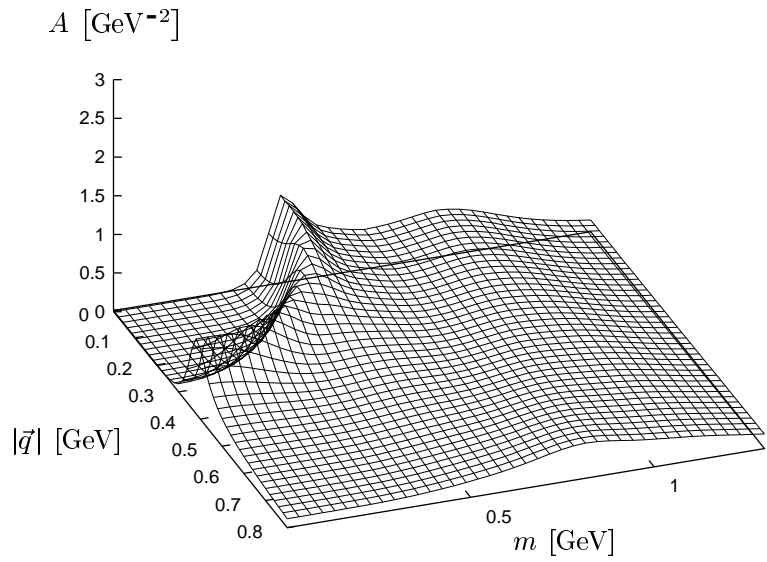,width=10cm}
 \caption{
\label{sc2}
Self-consistent averaged spectral function of the rho
 meson for $\rho_{_N} = 2 \rho_o$. }
\end{figure}

Fig.~\ref{sc2} shows $A$ calculated self-consistently for a density
$\rho_{_N} =2\rho_o$. One sees that the ridge coming from the
$\Delta(1232)$ becomes more prominent, since its contribution to the
self-energy increases (cf. Eqs. (\ref{beta}) and (\ref{lowdens})), 
but it is not broadened like  the other
resonances (see above).
At the same time, higher invariant masses are even more
depleted.  For higher $\vec{q}$ the rho branch is still visible,
 but it is much weaker than for $\rho_{_N} =\rho_o$
(Fig.~\ref{sca}).
The peak at $\vec{q}=0$ in the region of the $N(1520)$, which
acquires a large width in medium, 
is due to the fact that the rho-strength 
of the N(1520) depends very strongly on the mass of this resonance 
because of its 'subthreshold' character. Since this mass is lowered at 
the higher density, the rho-channel opens up and the strength increases 
significantly.

\begin{figure} 
\epsfig{file=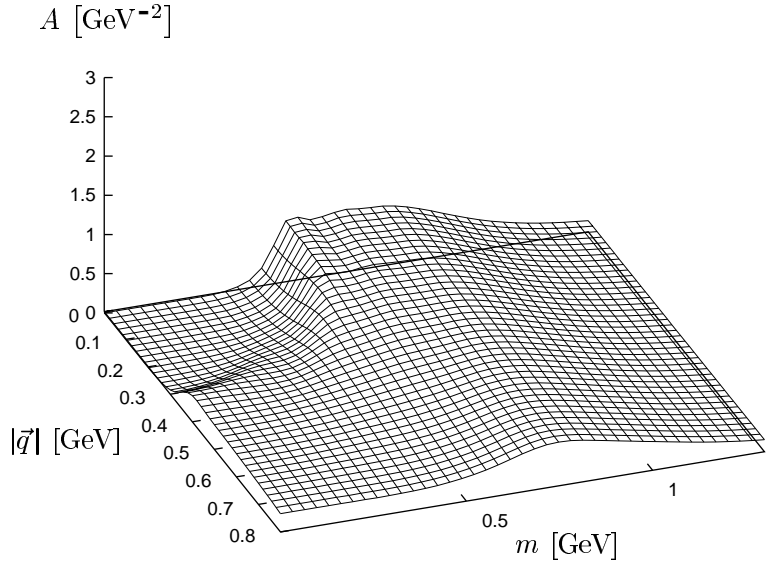,width=10cm}
\epsfig{file=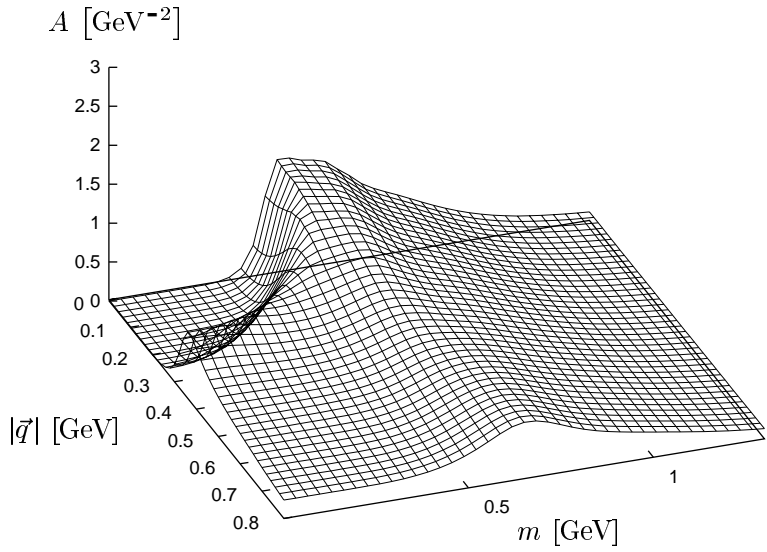,width=10cm}
 \caption{
\label{schwlm}
Self-consistent averaged spectral function of the rho
 meson for $\rho_{_N} = \rho_o$.
Upper part: using a vacuum rho-width of 300 MeV.
Lower part: using a rho mass reduced by 100 MeV.}
\end{figure}

We now explore the sensitivity of our results to additional effects on
the rho in medium, that are not yet included.  First, the modification
of the $\pi\pi$-loop has to be considered. In \cite{herrmann,chanfray}
the authors show, that this leads mainly to a broadening of the
rho. In the upper part of Fig.~\ref{schwlm} we therefore show $A$ as 
the result of a
self-consistent calculation employing twice the vacuum value for the
decay width of the rho. The effect is, that around $m=m_\rho$ the
strength is smeared out somewhat more, the rho-peak is less
pronounced.  The qualitative behavior of $A$ remains
unaltered.

Second, there still could be a possible change of the mass-parameter
in the propagator of the rho. This might be due to a Brown-Rho type
scaling, or due to $t$-channel interactions of the rho with the
nucleon as e.g. considered in \cite{friman0}. 
In the lower part of
Fig.~\ref{schwlm} we therefore show
results of a calculation employing a rho mass (see
Eq.~(\ref{fullspec})) that is lowered by 100 MeV.  
The rho peak at high $\vec{q}$ is shifted by 100 MeV, while at small
momentum a stronger shift to lower invariant masses is visible.

The negative parity  resonances, that we  assume to couple to a rho 
in an $s$-wave state,
lead to a strong effect in our
calculation. They can in principle also couple to a rho in a relative
d-wave, which would again vanish for $\vec{q}=0$. The main effect on
the rho spectral function, however, comes from the $N(1520)$ and in
the analysis
\cite{manley} the coupling of this resonance to the rho is clearly
determined to be $s$-wave. Besides, for the decay of a resonance
into a nucleon and a massive meson like the rho, the higher waves will
be the more suppressed, the lighter the resonance is.

        \section{Summary and Conclusions}
\label{concl}

We have calculated the transverse and longitudinal 
spectral functions of the rho meson in nuclear matter
in a model that includes the interaction of the rho with the medium
via resonance-hole loops. In extension to earlier works, we include all
resonances up to 1.9 GeV. We find that the so far neglected
$s$-wave resonances, especially
the $N(1520)$, have a very strong influence on the spectral
functions, especially for small three-momentum of the rho. 
We have performed a self-consistent
calculation by iterating the equation for the spectral functions. This
amounts to including terms of arbitrary order in the nuclear density.
We find that the self-consistent treatment of the rho
has a strong effect for small three-momentum of the rho, but leaves
both longitudinal and transverse spectral function
 more or less unchanged for $|\vec{q}\,| >0.5$
GeV. 
In the transverse channel
the rho-peak vanishes completely and a large amount of
strength is shifted to
smaller invariant masses. Besides a ridge coming from the
$\Delta(1232)$, no structure can be seen anymore; we find a continuum
of transverse excitations with the quantum numbers of the rho.
For the longitudinal spectral function we find a different 
behavior: Being equal to the transverse one for $\vec{q}=0$,
it shows a clear peak stemming from the rho meson for 
$|\vec{q}\,| >0.5$ GeV, which is reduced by a factor of two 
relative to the vacuum.

As a side result of our self-consistent calculation, we find that
simultaneously the widths of the resonances with rho-nucleon decay branches
are drastically enhanced in medium due to
the fact that the rho spectral function shows much strength at small
invariant masses; only the $\Delta(1232)$ is left unchanged. 
Since these are just those resonances 
that also have large electromagnetic couplings (vector meson dominance!) 
this behavior will contribute to the observed disappearance of the 
higher-lying nucleon resonances beyond the $\Delta(1232)$ in the total 
photoabsorption cross section on nuclei

We conclude, that the resonant nature of a transverse rho meson
vanishes completely in medium, leading to a strong enhancement at
small invariant masses. The higher the three momentum of the rho, the
broader is the distribution of the transverse strength, while in the
longitudinal case the rho branch is still visible for $|\vec{q}\,|
>0.5$.  Thus we find, that a simple scaling law of the rho-mass and
perhaps the width, that does not distinguish between longitudinal and
transverse polarization, cannot account for the effect of hadronic
interactions on the properties of the rho in medium.

\section{Acknowledgments}

One of the authors (U.M.) thanks the Institute for Nuclear Theory  
at the University of Washington for its hospitality and the U.S. 
Department of Energy for partial support during completion of this work.

\newpage

\begin{table}
\begin{tabular}{|r || l | r | r | r | r | c | c | l} \hline
           &                                    &              &               &          &             &    &         \\[-.35cm] 
           &$I(J^P)     $                       &$\Gamma_{\pi}$&$\Gamma_{\rho}$&$l_{\rho}$&$f_{R N\rho}$&$S_\Gamma$& $S_\Sigma$ \\ 
           &                                    &$[ MeV]        $&$[ MeV]       $&           &      &        & \\  \hline
  N(940)   &$\frac{1}{2}(\frac{1}{2}^{\,+})$    &0      &0      &1     & 7.7         &6    &   4        \\
  N(1520)  &$\frac{1}{2}(\frac{3}{2}^{\,-})$    &95     &25     &0     & 7.0         &1    & 8/3        \\
  N(1650)  &$\frac{1}{2}(\frac{1}{2}^{\,-})$    &135    &15     &0     & 0.9         &3    & 4        \\
  N(1680)  &$\frac{1}{2}(\frac{5}{2}^{\,+})$    &118    &12     &1     & 6.3         &3/5  & 6/5        \\
  N(1720)  &$\frac{1}{2}(\frac{3}{2}^{\,+})$    &50     &100    &1     & 7.8         &2    & 8/3        \\
$\Delta(1232)$&$\frac{3}{2}(\frac{3}{2}^{\,+})$&120     &0      &1     & 15.3        &2/3  & 16/9       \\
$\Delta(1620)$&$\frac{3}{2}(\frac{1}{2}^{\,-})$&130     &20     &0     & 2.5         &1    &  8/3       \\
$\Delta(1700)$&$\frac{3}{2}(\frac{3}{2}^{\,-})$&180     &120    &0     & 5.0         &1/3  & 16/9      \\
$\Delta(1905)$&$\frac{3}{2}(\frac{5}{2}^{\,+})$&140     &210    &1      & 12.2       &1/5  & 4/5 \\\hline 
\end{tabular}
\caption{
\label{listres}
List of the resonances included in this calculation together with the
parameters used. The third and fourth columns contain the values we
use for the pionic width of the resonances (see. Eq.~\ref{piwidth}))
and explanation thereafter) and for the rho-width. Columns five and
six show the angular momentum of the rho and the coupling constants
from Eq.~(\ref{couplings}). The last two columns contain the spin/isospin
factors occurring in Eqs.~(\ref{rdecay1},\ref{rdecay2}), (\ref{selfpos}) and
(\ref{selfneg}). }
\end{table}

\begin{table}
\begin{tabular}{| r || r | r | r | r | r |} \hline
& \multicolumn{4}{c|}{$\Gamma_\rho [MeV]$} \\ \hline
&Vacuum&Pauli-bl. &$\rho=\rho_o$&$\rho=2\rho_o$ \\ \hline 
        N(1520) &       25 &    7 &     240 &   200 \\ 
        N(1650)&        15 &    5 &      35 &   20 \\ 
        N(1680) &       12 &    8 &      90 &   60 \\ 
        N(1720) &       100 &   85 &    560 &   330 \\ 
        $\Delta(1620)$ &20 &    7 &      70 &   40 \\ 
        $\Delta(1700)$ &120 &   45 &    160 &   90 \\ 
        $\Delta(1905)$ &210 &   205 &   300 &   130 \\\hline
\end{tabular}
\caption{
\label{listwidths}
List of the free and in-medium rho-widths of the resonances on their
mass shell. The second column shows the free value we start from, the
third column is the free width corrected for Pauli-blocking. The
fourth and fifth column show the widths resulting from a self-consistent
calculation of the rho spectral function for two different densities.}
\end{table}

\clearpage

\begin{figure} 
\epsfig{file=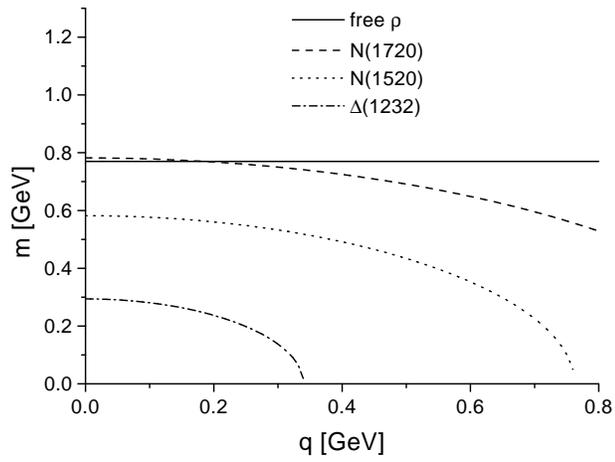,width=10cm}
 \caption{ 
\label{branches}
Free dispersion relations for the rho and the most prominent resonances.
For the resonance branches $m$ is the invariant mass a meson of given 
three-momentum $q$ must have in order to excite the respective resonance on
its mass shell.}
\end{figure}

\end{document}